# Computer Science Framework to Teach Community-Based Environmental Literacy and Data Literacy to Diverse Students

## Abstract

Clare Baek, Dana Saito-Stehberger, Sharin Jacob, Adam Nam, Mark Warschauer

University of California, Irvine

This study introduces an integrated curriculum designed to empower underrepresented students by combining environmental literacy, data literacy, and computer science. The framework promotes environmental awareness, data literacy, and civic engagement using a culturally sustaining approach. This integrated curriculum is embedded with resources to support language development, technology skills, and coding skills to accommodate the diverse needs of students. To evaluate the effectiveness of this curriculum, we conducted a pilot study in a 5th-grade special education classroom with multilingual Latinx students. During the pilot, students utilized Scratch, a block-based coding language, to create interactive projects that showcased locally collected data, which they used to communicate environmental challenges and propose solutions to community leaders. This approach allowed students to engage with environmental literacy at a deeper level, harnessing their creativity and community knowledge in the digital learning environment. Moreover, this curriculum equipped students with the skills to critically analyze political and socio-cultural factors impacting environmental sustainability. Students not only gained knowledge within the classroom but also applied their learning to address real environmental issues within their community. The results of the pilot study underscore the efficacy of this integrated approach.

**Introduction**

In today's technology-driven society, it has become increasingly essential to equip students with the knowledge and skills necessary for active civic engagement and addressing societal issues critically, utilizing computing as a tool. Among these skills, computer science proficiency stands out as an essential competency applicable across various disciplines, careers, and civic contexts (Bers, 2019; Stamatios, 2022; Resnick & Rusk, 2020). With the "Computer Science For All" initiative, there has been a growing implementation of computer science courses in K-12 education (Goode et al., 2018; Ladner & Israel, 2016). Additionally, as we face unprecedented climate challenges, environmental awareness and knowledge have become more crucial than ever. Efforts at international, national, and state levels have aimed to enhance environmental science education. For example, California has been committed to fostering environmental literacy throughout K-12 education, spearheaded by the California Department of Education (Lieberman, 2017). The United Nations Decade of Education for Sustainable Development similarly focuses on providing individuals with quality education to instill the behaviors necessary for sustainability (Biswas, 2020). Equipping students with environmental knowledge and awareness can inspire them to take responsible actions and commit to addressing environmental issues (Biswas, 2020). An integrated curriculum combining computer science and environmental science has the potential to enrich students' skills in both fields. This integration of computer science into other subjects not only enhances students' learning experiences but also cultivates their computing and computational thinking abilities (Fascali et al., 2018).

Furthermore, computer programming can serve as a powerful tool for creating innovative content that engages students in addressing societal issues and teaching scientific concepts (Yu et al., 2020). In this context, students are not merely learning to code but using coding to facilitate their learning of other academic skills, including critical thinking, problem-solving, and social skills (Popat & Starkey, 2019). Coding enables students to create projects that convey their stories and ideas, employing innovative features such as animation, audio, games, and images to emphasize their messages. For example, students have developed Scratch animations to address issues like racism, and teachers have utilized Scratch to teach the butterfly life cycle through animations (Resnick & Rusk, 2020). Teaching computer science within a context that is meaningful to students through projects and lessons that resonate with their interests can be a potent motivator for learning computer science skills (Cooper & Cunningham, 2010). Consequently, students not only learn how to code but also employ coding as a means to achieve their desired creative outcomes (Resnick & Rusk, 2020). This process also nurtures students' computational thinking skills, a set of cognitive processes involving algorithmic thinking to formulate and solve problems (Song et al., 2021). Creating projects through coding necessitates computational thinking, involving a systematic problem-solving process. Moreover, rather than solely focusing on learning programming techniques and syntax, applying programming skills to address real-world or STEM-related problems is a valuable approach for enhancing students' coding and computational thinking skills (Lin et al., 2018).

Data literacy has also become indispensable across academic disciplines and everyday life, playing a crucial role in domains ranging from economics to politics (Gebre, 2022; Shreiner, 2018). Data literacy is defined as the capacity to comprehend, analyze, interpret, communicate, and create data in a critical manner (Grillenberg & Romeike, 2018). Educating students on how to analyze and interpret data empowers them to make informed decisions, enhances problem-solving abilities across disciplines, and prepares them for active civic engagement (Gibson &

Mourad, 2018; Shreiner, 2019). Thus, initiating data literacy education as early as possible is crucial for equipping students with these essential skills.

Underrepresented students can particularly benefit from an integrated computer science curriculum that employs culturally sustaining pedagogy. Such an integrated curriculum creates a multisensory learning environment, as technology provides various modalities and sensory inputs, enhancing content comprehension for students with disabilities and English language learners (Anderson & Putman, 2020). For example, students can engage with content through verbal instructions, visuals, written text, audio, animation, and simulation. The diversity in learning channels supports students' understanding and retention of coding concepts and content knowledge. An integrated computer science curriculum, infused with culturally sustaining elements, can motivate students from diverse cultural and linguistic backgrounds by leveraging their rich cultural resources in the learning process. By connecting curriculum content to their own lives and experiences, this approach makes learning meaningful and fosters computing skills through innovative technology. An integrated computer science curriculum, featuring language and literacy scaffolding, alongside culturally sustaining lessons and projects where students collect data to address issues in their communities, holds significant value for culturally and linguistically diverse students. In such a context, students' cultural knowledge becomes assets as they engage in culturally relevant projects, collecting data that holds personal significance.

We have developed an integrated computer science curriculum infused with community-based environmental literacy. This curriculum incorporates language and literacy scaffolding, alongside culturally sustaining elements, to support diverse students. By involving students in projects that are relevant to their own communities and lives, we empower them to draw upon their cultural values. We piloted this curriculum framework in a 5th-grade classroom with Latinx students, all of whom had mild to moderate disabilities and were English language learners. The development of this curriculum framework and the pilot study represent the initial stages of a design-based implementation research project, focusing on teacher instruction, student learning, and issues arising in teaching practices as identified by practitioners, students, and researchers (DBIR; Penuel et al., 2011). The following research questions guide this study:

**Research question 1:** How can an integrated computer science curriculum framework be designed to promote environmental literacy and data literacy for students with diverse needs?

**Research question 2:** What were the affordances and challenges of the integrated curriculum in the learning and teaching of multilingual Latinx students with disabilities?

## Relevant Literature

### Environmental Literacy Curriculum

Environmental literacy refers to our ability to comprehend and interpret the overall health of environmental systems and take appropriate actions to maintain, restore, or enhance their well-being (Roth, 1992). The aim of environmental education is to equip individuals with the knowledge and skills needed for successful, productive lives and responsible citizenship (Roth, 1992). Environmental literacy plays a foundational role in instilling children's values and commitment to environmentally friendly practices (Biswas, 2020). Thus, initiating environmental literacy education at an early stage in K-12 education is crucial.

Existing studies on environmental science curriculum for elementary school students have highlighted the effectiveness of community-based environmental lessons. For example, Cronin-Jones (2000) found that students who received environmental lessons through outdoor exploration of their schoolyard exhibited significantly greater gains in environmental science content knowledge compared to those in traditional classroom settings. The outdoor schoolyard group engaged in activities such as field observations of the schoolyard plants and animals, in contrast to traditional slide-based instruction. Upadhyay and DeFranco (2008) discovered that students who received environmental science instruction through the connection of lessons to real-life situations and community issues, rather than explicit instruction, demonstrated better retention of environmental knowledge. In a more recent study, Cheng et al. (2019) used technology to explore students' learning experiences in environmental science. Cheng et al. (2019) compared the environmental attitudes and environmental science content attainment of elementary school students who received traditional instruction via a mobile device app with those who engaged in situated learning using the same mobile device app. Students in the situated learning group observed, collected, analyzed data, and adjusted an aquaponics environment created in the classroom while utilizing the app, whereas the conventional learning group used the app to learn content without hands-on experience. The situated learning group achieved significantly higher learning outcomes, demonstrated improved environmental attitudes, and displayed enhanced problem-solving skills.

**Computer Science Curriculum**

A computer science curriculum not only can enhance students' coding skills and computational thinking but can also nurture their identity and literacy skills. A computer science curriculum that centers on culturally responsive relevant pedagogy provides a platform for multilingual Latinx students to express their identities by creating Scratch projects about their life stories, families, and communities (Jacob et al., 2020). As students learn coding, which is a form of language, they refine their language and literacy skills (Bers, 2019). However, teaching coding concepts and computational thinking to English language learners can be challenging.

Effective instructional strategies for teaching computational thinking to English language learners require robust linguistic scaffolding to guide their understanding of discipline-specific discourse structures and vocabulary, foster academic language and content knowledge, and integrate instructional practices with culturally responsive materials (Jacob et al., 2018). Research by Weng and Wong (2017) demonstrated that introducing computational thinking into English dialogue learning via Scratch improved elementary school students' motivation to study English. Parsazadeh et al. (2021) also showed that a Scratch learning activity integrating computational thinking into English language learning was effective in enhancing elementary school students' English language proficiency.

Students with disabilities may encounter barriers when accessing and engaging with computer science curricula. However, with appropriate support, they can learn programming just as effectively as their peers without disabilities (Adebayo, 2022; Israel et al., 2015; Ladner & Stefik, 2017). For example, Ratcliff and Anderson (2011) found that engaging in computer programming projects aligned with students' interests led to improved attitudes, increased enjoyment, and a greater sense of ownership of their learning. Taylor et al. (2017) similarly demonstrated that students with Down Syndrome can successfully learn block-based coding. A computer programming environment can enhance the learning experience and outcomes for students with disabilities in areas beyond coding. The collaborative nature of the computing process, focused on creativity and problem-solving, enhances peer interactions and social engagement for students with autism, who may typically struggle with social interactions (Gribble et al., 2017; Munoz et al., 2018). Additionally, coding in an interactive block-based programming environment can effectively improve writing skills for students with learning disabilities (Thompson, 2016).

**Data Literacy Curriculum**

Traditionally, data literacy has been considered a skill for older students. However, recent efforts have aimed to promote data science education in primary and secondary education (Jiang et al., 2022). Younger students can develop data literacy and enrich their content knowledge through collecting, analyzing, creating, and presenting data. Engaging students in meaningful inquiries about relevant and significant data, such as ecological systems in their own environment or neighborhood, can foster their interest and involvement in data-related inquiries. As exemplified by Cheng et al. (2019), elementary school students can collect, analyze, and present data from ecological systems using mobile applications. Wolff et al. (2019) demonstrated that students aged 10 to 18 can utilize interactive maps to explore their own neighborhoods and investigate factors influencing the suitability of houses for solar panels. Further research on younger students' data literacy learning and strategies to integrate data literacy into other subjects is timely and necessary.

<div align="center">

**Theoretical Perspective**

</div>

The integrated curriculum is guided by culturally sustaining pedagogy, emphasizing the active engagement of students in their own communities to collect data on pertinent local issues. This approach not only fosters a deeper connection with the environment but also cultivates an awareness of influential social structures. Moreover, it empowers students to share their awareness actively with others. When culturally sustaining pedagogy is integrated into science lessons, it heightens students' consciousness of community and environmental concerns while nurturing their environmental literacy (Upadhyay et al., 2020). By engaging in discourse about local environmental issues, students gain the ability to critically examine the economic, social, and cultural factors affecting their community, bridging the gap between science content and real-life experiences (Freire, 1998; Morales-Doyle, 2017; Seider et al., 2017).

A culturally sustaining curriculum creates an inclusive learning environment for culturally and linguistically diverse students. Centered on linguistic and cultural pluralism, such a curriculum embraces students' diverse cultural backgrounds and identities (Paris & Alim, 2014). A culturally sustaining curriculum can hold particular significance for Latinx students who value community and culture.

## Integrated Curriculum Framework

Our current curriculum places a strong emphasis on students' cultural backgrounds and their understanding of their local community. To develop the environmental science content, we followed the Environment as an Integrating Context (EIC) model and California's Environmental Principles and Concepts (EP&Cs). The EIC model underscores the relationship between natural systems and human society, highlighting the impact of human society on natural systems. Furthermore, it delves into how decision-making processes influence choices concerning resources and natural systems, guiding students to explore real-world issues and knowledge as active citizens (Lieberman, 2013). We collaborated with the developer of the EIC model to ensure that the curriculum's content aligns with the EIC model's principles. The EIC model promotes a language-rich and culturally sustaining learning environment conducive to computing by engaging students in collaborative community-based learning, offering both cognitive and hands-on learning experiences (Lieberman, 2013). Rather than discussing distant topics like polar ice caps, students delve into their own local environmental issues, enhancing their motivation and sense of agency to effect tangible change.

The integrated curriculum framework that we devised, illustrated in Figure 1, comprises six sequential components: driving questions, tools, data, product, skills, and knowledge. The driving questions form the basis for what students wish to explore and how they intend to find answers. To ensure cultural sustainability, these driving questions should resonate with students' own lives, focusing on environmental issues within their community (e.g., school, neighborhood), family heritage, language, and culture. By utilizing computational tools, students gather and analyze data relevant to addressing these driving questions (Figure 2). Based on the collected data and their analysis, students create products that depict their findings, highlight issues, and propose potential solutions. Through this process, students develop problem-solving skills, computational thinking abilities, and scientific communication proficiency. Ultimately, these skills culminate in knowledge, including environmental literacy, data literacy, and civic engagement.

The lessons involved students using Scratch to create projects based on the environmental concepts they learned. These lessons were integrated with literacy resources, including word banks and sentence frames. Initially, students were introduced to the broader context of natural and human systems (Figure 3). Subsequently, they explored the interactions between these systems. As the lessons progressed, students honed in on their own environment, guided by driving questions addressing the natural and human systems in their school, their interactions, and ultimately, the main question explored in the final project: "What environmental issues exist in my school, and how can we mitigate them?"

The culminating project, titled "Imagining Change," guided students in identifying environmental issues within their school environment and documenting these problems by taking pictures of the problems. Students then devised solutions to address these identified environmental issues and created projects that showcased "before" and "after" representations of their proposed changes, supplemented by persuasive messages in either text or audio format, advocating for their environmental solutions to school administrators. For instance, a student identified the lack of greenery in the school quad area as a problem, attributing it to heavy foot traffic and a lack of planted vegetation. As a solution, the student proposed planting more sustainable plants. Throughout this process, the students employed scientific language, incorporating environmental literacy vocabulary, sentence structures, and sentence frames to elucidate environmental phenomena. The students subsequently presented their "Imagining Change" projects to the school vice principal to advocate for improvements to the school environment.

To provide optimal support for students with disabilities and English language learners, a set of resources and prerequisite lessons were integrated, as depicted in Figure 1. Before using data collection tools, students were equipped with visual cues, a language reference sheet, and models. These resources were designed to assist students in effectively collecting data, including instructional models on how to use citizen science apps for investigating plant types around the school, a language reference sheet containing environmental literacy vocabulary (e.g., phenomenon, natural system), and visual cues with different colors representing environmental phenomena (e.g., less green space, more green space on a school map). Additionally, we ensured that students possessed the necessary skills before employing the collected data to create their projects. For example, students received training in making presentations using PowerPoint, where they created slideshows on topics of their choice (prerequisite ICT skills) in preparation of the lesson on creating PowerPoint to present the pictures they collected to class. They also learned how to edit pictures they had taken by cropping and adjusting sizes (prerequisite ICT skills). To support students' written and oral communication, we provided example sentence structures for composing persuasive texts. Moreover, a "cheat sheet" containing Scratch blocks and functions was made available to assist students during the coding process in Scratch. An example illustrating how we followed the curriculum framework is presented in Table 1.

To address the data literacy component of the curriculum, we adapted the stages of data competency proposed by Grillenberger and Romeike (2018). These stages encompassed planning for data collection, data collection, data cleaning, data analysis, and data presentation for the final project, "Imagining Change." During the planning stage, the class collectively established criteria for the data they intended to collect to address campus environmental issues (e.g., pictures and information about plants on campus). Subsequently, students engaged in data collection by taking pictures of the identified environmental problems on campus and collecting information about the types of plants growing there. In the data cleaning stage, students assessed the data quality and refined the collected data. For example, students selected useful data by evaluating picture quality and relevance in depicting the environmental problems they aimed to address. Students cleaned the selected data by cropping pictures to enhance visibility where applicable. In the data analysis step, students interpreted the data to critically assess the environmental problems represented in the pictures and information they had gathered, examining their solvability and potential solutions. Finally, during the data presentation stage, students conveyed their data analysis results using visual and interactive presentations in Scratch, incorporating persuasive text delivered through audio recordings of their voices or in text form, based on their preferred modality.

**Pilot Study**

**Methods**

We conducted the integrated curriculum framework in a classroom setting to explore the learning and teaching experiences and outcomes for diverse students. We chose an exploratory case study approach because we intended to investigate our research questions without specific hypotheses. The case study methodology was suitable as it allowed us to gain insights into the phenomena as they unfolded during the integrated curriculum's implementation in this specific classroom. This, in turn, provided valuable guidance for refining the framework's design and implementation (Yin, 2008; Yin, 2011). The case in this study was the implementation of the integrated computer science curriculum with community-based environmental literacy. The unit of analysis encompassed the entire class ecosystem, including the teacher, paraeducator, students, and learning activities.

This case study was conducted in a 5th-grade mild-to-moderate special day class in California with 12 students, consisting of Latinx students classified as English language learners. Before conducting the study, parent permission letters and student information letters were sent out in both English and Spanish. The participants had the option to opt out of the study, which was clearly communicated in the permission letters and information letters. The curriculum was co-designed by the first and second authors in collaboration with the classroom teacher, Ms. Linda, a special education teacher with 12 years of teaching experience. Ms. Linda, also an environmental lead teacher for the district, was responsible for co-designing the environmental literacy curriculum with the developers of the California environmental literacy curriculum and training teachers in the district. Ms. Linda did not possess prior coding experience or experience in computer science teaching. To ensure ease of implementation for teachers without coding experience, we designed the curriculum to include slides for each lesson with guided instructions, comprehensive lesson plans, Scratch project examples for students, and additional resources on Scratch skills.

While environmental literacy designed with the EIC model had been widely implemented in the school district, it had not been previously integrated with a computer science curriculum. The objective of this project, in partnership with the district, was to develop an integrated curriculum encompassing computer science for district-wide implementation across elementary schools. Ms. Linda delivered the curriculum once a week during the 2022-2023 school year. Throughout the study, the first and second authors visited the classroom weekly as participant observers.

**Data Collection and Analysis**

The collected data included co-design meeting notes with the classroom teacher, audio recordings of classroom observations, audio recordings of informal student interviews, audio recordings of semi-structured teacher interviews, field notes, and student artifacts. Student artifacts consisted of their literacy scaffolding sheets, student-collected data (e.g., pictures taken by students, campus maps indicating green spaces), and Scratch projects. Audio recordings of classroom observations and interviews were transcribed using transcription software for subsequent analysis.

We conducted a thematic analysis using an inductive approach to examine the transcribed audio data, addressing our research question regarding the affordances and challenges of the integrated curriculum for teaching and learning in diverse student populations. The first and fourth authors collaboratively followed the steps of thematic analysis, commencing with iterative readings of the transcriptions and the generation of initial ideas through multiple cycles (Braun & Clarke, 2006). Initial codes were subsequently generated, with excerpts from the transcriptions relevant to each code being compiled. These compiled codes were then associated with potential themes, and the research team reviewed the themes to ensure that the coded excerpts corresponded appropriately with the themes. The first and fourth authors worked collaboratively to define the themes after iteratively refining them. Excerpts were re-evaluated to identify any additional codes that aligned with the established themes and were added to the relevant themes when applicable. To ensure consistency across the data, we triangulated data from multiple sources, including interview transcripts, observation transcripts, and student artifacts.

## Results of the Pilot Study

We discovered the unique affordances of the integrated curriculum. Students showed heightened engagement with the curriculum and engaged with environmental concepts in a deeper way than the traditional environmental literacy curriculum. Students gained confidence as they created projects that were relevant to them by collecting their own data and presenting problems and solutions through the collected data while expressing their creativity through coding. Students promoted their environmental literacy and environmental awareness as they connected classroom lessons to addressing issues in their own community (Figure 4). Students were able to apply the knowledge they acquired in the classroom to notice environmental issues outside the classroom and engaged in discussion. Students honed their everyday language skills as they engaged in coding. There were challenges to implementing the integrated curriculum, including a lack of time to complete Scratch projects and managing data.

### Theme 1: Environmental Literacy Engrained in Students Through Integrated Curriculum

Ms. Linda shared that with the integrated approach, students were able to learn the environmental concepts much better than with the traditional environmental literacy curricula.

Students completed the lessons and projects using an innovative digital environment, computer science, which created a fun learning environment with multiple modes of learning.

> With the computer science component of it…it's like a double dose basically…because we would do the slides, we would do videos, then we would do the coding…and we don't usually get that opportunity. So it's like…they are getting it different ways.

The culturally sustaining pedagogy was particularly valuable for students' engagement in the projects and promoted their confidence. Ms. Linda stated that because the projects were related to students' daily lives and students created projects in Scratch by integrating their own creativity, it made the lessons more valuable to the students. She emphasized that the projects were "really theirs."

> We were able to merge computer science with the environmental literacy units….it turned out to be, um, a little bit more personal for the students…they made it their own…because…it's their environment, it's their community.

Even when the students were outside the classroom setting, Ms. Linda witnessed them paying attention to their surroundings and addressing the environmental issues. Ms. Linda stated that the student had developed an understanding of the importance of the environmental component:

> They [students] are still saying like, oh there's too much trash here…the trees are dying…I feel like it's been engrained in their minds now.

Ms. Linda added that she thinks the students will take the lessons they learned through this curriculum forever, "I feel like they will take this with them forever."

**Theme 2: Students Leading Culturally Relevant Discussions**

During class, students took the initiative in leading discussions by sharing stories and knowledge about culturally relevant environmental topics. For instance, one student discussed the presence of mango trees at his uncle's house while the class explored the topic of trees. This led to a broader discussion about where the uncle lived, the climatic conditions required for mango trees to thrive, and the steps the uncle took to plant and cultivate the trees.

Another instance involved students initiating a class discussion during an environmental literacy lesson about their destroyed class garden. The students discussed how the garden they had created was destroyed by school staff, who had used it for a school event without obtaining permission from the class. This discussion expanded to encompass the importance of seeking and obtaining permission to protect an environmental system, eventually leading to a discussion of more extensive systemic environmental issues and the influence of political and socio-cultural factors on decisions affecting the environment.

**Theme 3: Students Practicing Literacy Through Coding**

Another emerging theme highlighted how the coding process promoted literacy skills. As students debugged their code to identify errors that prevented it from functioning correctly, they sought assistance from the teacher or other students. This process required clear explanations of the steps and sentence structures, contributing to students' literacy skills.

Additionally, students' vocabulary was activated as they worked with Scratch blocks with various functions. For example, a student explained that their code wasn't working because the blocks were not connected, and when the teacher inquired about this, the student used the synonym "detached." Interacting with Scratch blocks with different functions enabled students to engage with literacy by describing their coding needs and preferences. For example, one student communicated that he wanted the blocks with a "looks" feature so he could make the characters "appear" one by one.

**Theme 4: Challenges Experienced Including Lack of Time and Managing Data**

Students encountered several challenges while working with the integrated curriculum. The inclusion of multiple components, such as creating projects in Scratch, debugging errors, uploading data, and writing or recording persuasive messages, sometimes led to insufficient time to complete projects as envisioned. Some students shared that they had to choose alternative ideas due to time constraints, preventing them from fully realizing their original concepts. These challenges underscore the importance of allowing students opportunities to problem-solve and refine their projects to meet their intended goals.

Data management presented another challenge. Many students struggled with handling files, including downloading images from the web and saving them in easily accessible locations. To address this, students required step-by-step guidance from teachers. With repeated practice and support, most students eventually gained the skills to manage data independently.

<div align="center">

**Discussion**

</div>

The pilot study illustrates the effectiveness of an integrated curriculum delivered through a culturally sustaining approach. Incorporating coding to address real-world environmental problems specific to students' communities demonstrated promising results in enhancing the motivation and confidence of underrepresented students (Crooks et al., 2015). This approach is particularly beneficial for Latinx students, as it allows them to leverage their cultural knowledge and enrich classroom discussions (Luna & Martinez, 2013).

Furthermore, students with disabilities and English language learners successfully completed the projects with appropriate support and scaffolding. The block-based coding environment proved effective in engaging students with disabilities, enabling them to learn coding without heavy syntax while sharing their creativity and gaining environmental literacy simultaneously. Incorporating technology, such as computer programming, as a learning tool enabled students with disabilities to express their ideas in ways that might have been challenging using traditional methods. The support provided, including resources and prerequisite skills, effectively enhanced the learning process for students with disabilities and those requiring additional literacy support. Context-based projects in the coding environment contributed to the improvement of language and literacy skills among English language learners. In data literacy education, contextual relevance and meaningfulness of the data are important as well (Gebre, 2022). By presenting data that resonates with students' experiences, they became more motivated and actively engaged in the learning process.

The curriculum also promoted awareness of environmental health and justice. As students began noticing environmental issues in their communities through the curriculum, they developed skills for civic engagement. These discussions not only enhanced their critical environmental literacy but also turned them into active digital creators for civic engagement. The curriculum framework displayed adaptability and applicability across various subjects beyond environmental literacy.

The integrated curriculum design and implementation could benefit from several improvements. In the future, we aim to embed extra time in the lessons dedicated to students problem-solving and finishing what they could not finish in previous class time. Another challenge that the students experienced was with data management. During the prerequisite skills training session in the future, we aim to emphasize data management and guide the students on downloading files and saving them to the right place to find them. Also, greater emphasis should be placed on data ethics and the practice of crediting sources from the beginning of the curriculum, enabling students to develop this essential skill throughout the class (Carlson et al., 2011).

**Funding**

This work was supported by the National Science Foundation under Grant Number 1923136.

The conduct of this research has been approved by the University of California, Irvine Institutional Review Board #20173675.

**Figures**

Figure 1

*Curriculum Framework*

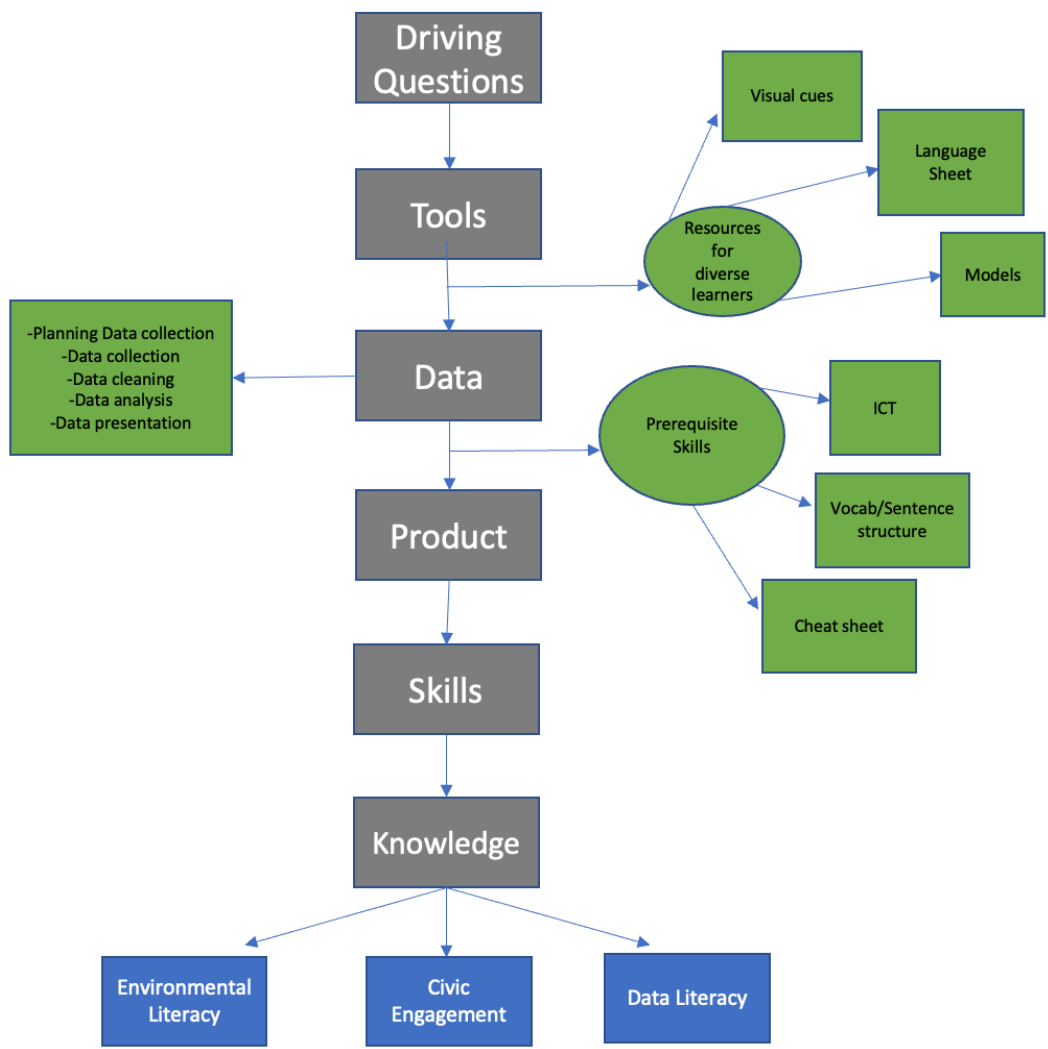

Figure 2

*Examples of Collected Data on Environmental Problems*

**They need more flowers**

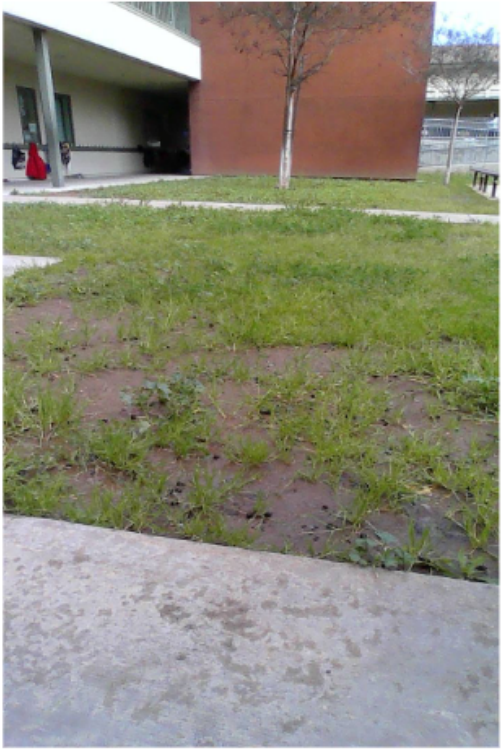

**We need more flowers and plants**

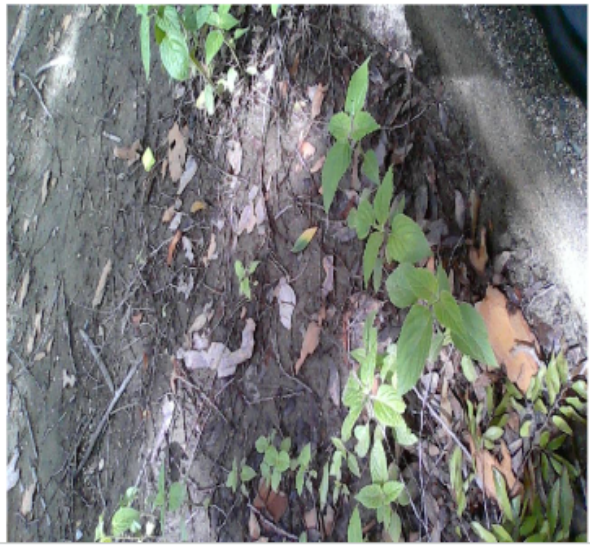

Figure 3

*Example of Students Learning Literacy*

Review your notes about trees, then answer the questions by filling in the blanks below.

1. What is one way that trees benefit people?

Trees give people _Oxygen_  → ?

2. What is one reason that trees are disappearing?

_Climate_ is causing trees to disappear. ?

3. What is one thing we can do to protect trees?

We can protect trees by _Telling friend how important trees are_ ?

Circle a picture to represent the 1, 2, and 3 statements above. Draw arrows to represent the relationships to the tree.

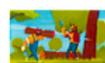

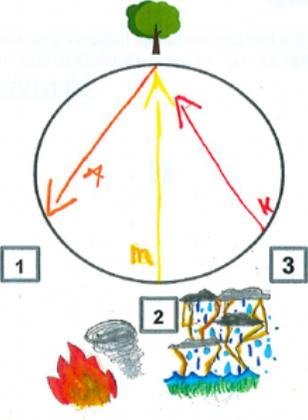

Figure 4

*Examples of the Imagining Change Project*

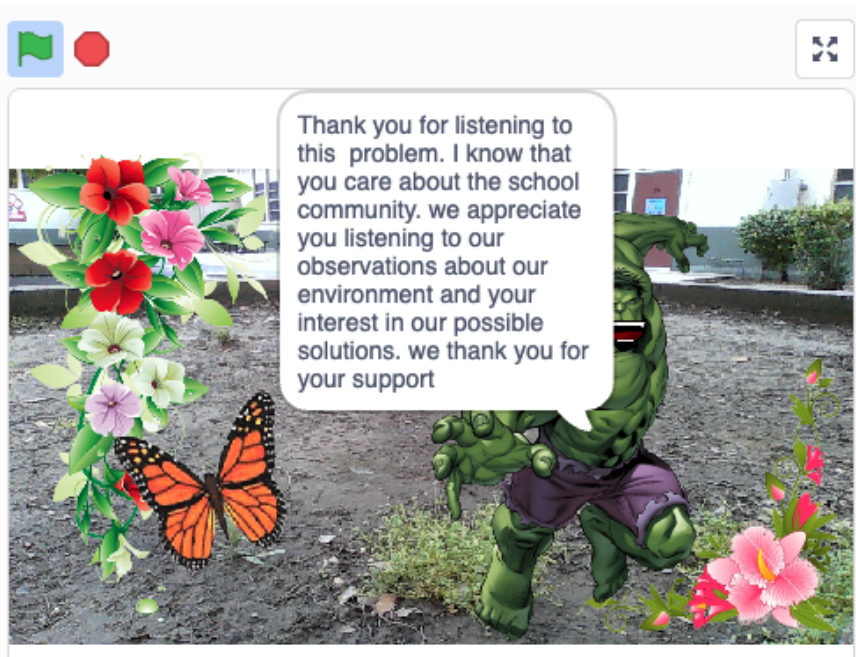

**Table**

Table 1

*Example of the Curriculum Framework Implementation*

| Curriculum Components | Examples |
|---|---|
| **Driving Questions** | • What do you want to know about your own environment?<br>• How can you find out what you want to know?<br>• What changes do you want to make?<br>• How can you communicate the changes? |
| **Tools** | • Environmental literacy graphic organizer<br>• Citizen science tool (App)<br>• Chromebook (to take pictures) |
| **Data** | • Pictures of the school environment<br>• Pictures from the web<br>• Information about the plants and flowers of the school gathered from Apps<br>• Criterion and constraints for engineering their solution |
| **Product** | • Scratch projects<br>• Presentations on environmental pictures collected<br>• Persuasive writing on environmental issues |
| **Skills** | • Computer science skills<br>• Computational thinking skills<br>• Data literacy<br>• Language/literacy |